\documentclass{aa}
\usepackage{psfig,amsmath}

\def\less{\,\raise 0.6ex\hbox{$<$}\kern -0.75em\lower 0.47ex \hbox{$\sim$}\,}
\def\more{\,\raise 0.6ex\hbox{$>$}\kern -0.75em\lower 0.47ex \hbox{$\sim$}\,}

\def\arcmin{\hbox{$^\prime$}}
\def\arcsec{\hbox{$^{\prime\prime}$}}

\begin{document}

\title{The chemistry of compact planetary nebulae
\thanks{Based on observations carried out with the IRAM 30m telescope. 
IRAM is supported by INSU/CNRS (France), MPG (Germany) and IGN (Spain).}
}

\author{
     E. Josselin  \inst{1}
\and R. Bachiller \inst{2}
}

\offprints{josselin@graal.univ-montp2.fr}

\institute{
GRAAL--CC72,  UMR 5024--ISTEEM, CNRS/Univ. Montpellier II, 
F-34095 Montpellier Cedex, France
\and
Observatorio Astron\'omico Nacional (OAN), IGN, Apartado 1143, 
28800 Alcal\'a de Henares, Spain
}

\titlerunning{The chemistry of compact planetary nebulae}

\date{Received / Accepted }

\abstract{  We report high-sensitivity millimetre observations of
several molecular species ($^{13}$CO, HCN, HNC, CN, HCO$^+$ and N$_2$H$^+$) 
in a sample of compact planetary nebulae. Some species such as 
HCO$^+$ and CN are particularly  abundant compared to envelopes around AGB
stars or even interstellar  clouds. We have estimated the following
average values for the column densities  ratios: CN/HCN$\sim$2.6,
HCO$^+$/HCN$\sim$0.5, and  HNC/HCN$\sim$0.4. Thus, the
chemical composition of the molecular envelopes in these compact PNe
appears somewhat intermediate between the composition of proto-PNe
(such as CRL\,2688 or CRL\,618) and well evolved PNe (such as the
Ring, M4--9, or the Helix). From observations of the CO isotopomers,
we have estimated that the $^{12}$C/$^{13}$C ratio is in the  range  $10 
\less ^{12}C/^{13}C \less 40$. These values  are below those expected
from standard asymptotic giant branch models  and suggest non-standard
mixing processes.  The observed molecular abundances are compared to
very recent modelling work, and we conclude that the observations are
well explained,  in general terms, by time-dependent gas-phase
chemical models in which the ionization rate is enhanced by several
orders of magnitude with respect to the average interstellar
value. Thus, our observations confirm that the  chemistry in the
neutral shells of PNe is essentially governed by the high energy
radiation from the hot central stars.  The complexity of the chemical
processes is  increased by  numerous factors linked to the properties
of the central star and  the geometry and degree of clumpiness of the
envelope.  Several aspects of the PN chemistry that remains to be
understood are discussed within the frame of the available chemical
models.
}

\maketitle

\section{Introduction}
While planetary nebulae (PNe) are primarily identified from their optical 
emission arising in the ionized gas, they often  contain a substantial 
amount of neutral gas which is a more direct relic of the mass 
lost on the asymptotic giant branch (AGB). 
CO and H$_2$ surveys of PNe have revealed 
that a large fraction of the gas in massive PNe is in molecular form 
(Huggins et al. 1996, Kastner et al. 1996). 
This molecular component may be dominant during a large 
part of the evolution of the nebulae, and thus plays a 
prominent role in the mass distribution and the shaping of PNe. 

The molecular gas in PNe is affected by the 
large radiation fields from the central stars which are expected to 
cause a dramatic effect on the chemical composition. 
The studies of the chemistry 
of the neutral gas in PNe have been very scarce so far.
Cox et al. (1992) observed the evolved PNe NGC 6072 and IC 4406
at millimetre wavelengths, and detected HCN, HNC, CN and HCO$^+$ 
in both objects. More recently, Bachiller et al. (1997a) 
made a more complete study, covering an evolutionary 
sequence. If we except the two PPNe of their sample, five 
PNe were searched and detected over this large variety of 
molecular species: the young PN NGC 7027 and the evolved ones 
NGC 6720 (the Ring), M 4-9, NGC 6781 and NGC 7293 (the Helix). 
The molecular content of proto-PNe has also been investigated by observing
some selected objects such as CRL 618 (Bujarrabal et al. 1988 ; 
Cernicharo et al. 2001), and OH 231.8+4.8 (S\'anchez Contreras et al. 1997).
It clearly appears from these studies that the chemical evolution
is dominated by photodissociation, ion/radical-molecular 
reactions, and shocks. In particular, the survival of some
molecules at well advanced stages of the evolution 
is made possible thanks to the shielding of the radiation, 
which is directly connected with the presence of inhomogeneities 
in the envelope. Very dense clumps, directly observed as  cometary globules
in the nearest PNe such as the Helix 
(Huggins et al. 1992, 2002), provide an environment which is well suited
to the protection of the molecules and of its on-going chemistry. 
The chemistry of the PNe neutral envelopes has been modelled by 
several authors. The first work was done by Black (1978) who 
studied the neutral-ionized transition zone in an homogeneous nebula.
Howe et al. (1992, 1994) modelled the evolution of a dense clump similar
to those of the Helix nebula. More recently, Hasegawa et al. (2000) 
and Hasegawa \& Kwok (2001) reported detailed models of the neutral 
envelope of NGC\,7027, and Ali et al. (2001) constructed
time-dependent gas-phase models which reproduce well
the molecular abundances in evolved PNe such as NGC\,6781, M4-9 and 
the Helix,  observed by Bachiller et al. (1997a). 

With the remarkable exception of
NGC\,7027, the previous chemical studies carried out so far 
have been mainly devoted either to rather evolved 
PNe (such as the Ring or the Helix) or to proto-PNe (such as CRL\,618
or OH\,231.8+4.8). 
We report here on high-sensitivity observations of molecular lines 
in a sample of 7 compact, young or intermediate-aged PNe, which improve 
our knowledge of the chemistry during the nebular development. 
The sample and the observations  
are presented in Sect. 2. An account of the results is given 
in Sect. 3. These results are discussed in Sect. 4 making  
particular emphasis on the chemistry of N-bearing compounds, on the
formation of molecular ions, and on the importance of the clumpiness. 

\section{Sample and observations}
\subsection{PNe sample}

\begin{table*}
\caption[ ]{Properties of the sample of PNe. }
\label{prop}
\begin{flushleft}
\begin{tabular}{lcccccccc}
\hline\noalign{\smallskip}
Name & PN G & D$_{\rm K}$ &    size     &   V$_{\rm lsr}$   &   V$_{\rm exp}$   & I$_{^{12}{\rm CO}(2-1)}$ & M$_{\rm m}$                 & M$_{\rm m}$/M$_{\rm i}$ \\
     &      & (kpc) & ($\arcsec$) & (km s$^{-1}$) & (km s$^{-1}$) &   (K km s$^{-1}$)  & (10$^{-2}$ M$_\odot$) &             \\
\noalign{\smallskip}
\hline\noalign{\smallskip}
BV 5-1   & 119.3+00.3  & 5.5 & 20 & -73 & 10 & ~~9.7 & ~2.8 & ~0.3 \\
K 3-94   & 142.1+03.4  & 5.5 & ~7 & -69 & 16 & ~14.1 & ~3.4 & ~0.2 \\
M 1-13   & 232.4--01.8 & 2.0 & 10 & ~27 & 18 & ~22.2 & 25.4 & ~1.8 \\
M 1-17   & 228.8+05.3  & 2.5 & ~3 & ~28 & 39 & ~66.2 & ~6.1 & 14.0 \\ 
K 3-24   & 048.7+02.3  & 3.5 & ~6 & ~44 & 24 & ~23.2 & ~2.1 & ~0.2 \\
IC 5117  & 089.8--05.1 & 3.0 & ~1 & -11 & 14 & ~19.6 & 24.5 & ~2.1 \\
NGC 7027 & 084.9--03.4 & 0.7 & 14 & ~26 & 23 & 278.0 & 22.0 & ~9.4 \\
\noalign{\smallskip}
\hline
\noalign{\smallskip}
\end{tabular}
\end{flushleft}
\end{table*}

Our sample is made of 6 compact PNe covering a wide range of evolutionary 
status and whose molecular chemistry has not been studied up to now. 
These are BV 5--1 (see Josselin \& Bachiller 2001 for its CO content), 
K 3--94, K 3--24 (Josselin et al. 2000), M 1--13, M1--17 and IC 5117 
(Huggins et al. 1996 and references therein). 
Although it was already observed by Bachiller et al. (1997a), 
as representative of young PNe, we also observed NGC\,7027 for comparison
and calibration purposes.
Table \ref{prop} lists the main properties of the observed 
nebulae: the most usual name (Col. 1), standard galactic name (2), 
kinematic distance - based on the 
simple approximation for the velocity structure of the Galactic disk 
of Burton (1974) - (3), optical size of the ionized nebula (4), radial 
and expansion velocities measured in the CO(2--1) line (5, 6), 
integrated intensity of the CO(2--1) line (7), 
mass of the molecular envelope (8), and ratio M$_{\rm m}$/M$_{\rm i}$ 
between the molecular mass and the ionized mass (9), 
which is a good indicator of the evolutionary stage, almost independent of 
the distance (Huggins et al. 1996). 
Concerning the distance, we have preferred the kinematic estimate over 
the statistical estimate, since we believe that the former is more
precise in these cases (see Josselin et al. 2000 for details). 
Nevertheless, for  NGC~7027, the  best studied object in the sample,
in accordance with previous works,
we have assumed the standard estimate of 0.7\,kpc 
(see e.g. Huggins et al. 1996).In the case of M 1-13, M 1-17 
and IC 5117 the values of  M$_{\rm m}$ and M$_{\rm m}$/M$_{\rm i}$ have 
been re-scaled from those given by Huggins et al. (1996) according to 
our choice of distance (M$_{\rm m}$~$\propto$~distance$^2$  and 
M$_{\rm m}$/M$_{\rm i}$~$\propto$~distance$^{-1/2}$)\footnote{the ionized 
mass is 
derived from the $\lambda$6~cm emission and the angular nebular size, 
resulting in a distance$^{5/2}$ dependance, as shown by Gathier (1987).}. 

\subsection{Observations}
The observations were carried out with the 
IRAM 30-m radiotelescope on Pico Veleta (near Granada, Spain)
in July 1999. The search 
for molecular emission was made simultaneously in four lines. 
We used SIS receivers operating in the bands around $\lambda$ 3 mm 
and $\lambda$ 1 mm with typical system temperatures 
of 150 and 300 K, respectively. 
The half-power beam width of the telescope is 22$\arcsec$ and 10$\arcsec$  
at 3 and 1 mm respectively. The spectrometers were 256$\times$1 MHz filter 
banks providing a velocity resolution of 2.6 and 1.3 km/s at 3 and 1mm, 
respectively. The observations were done by wobbling the secondary mirror 
to a distance of 2$\arcmin$ from the source. 
The pointing accuracy, checked every hour, was within 3$\arcsec$. Calibration 
was achieved with a chopper wheel. The intensities given hereafter are in 
units of main beam brightness temperature. 

\section{Results}
\subsection{Spectra}
\begin{figure*}[ht]
\centerline{\psfig{angle=-90,figure=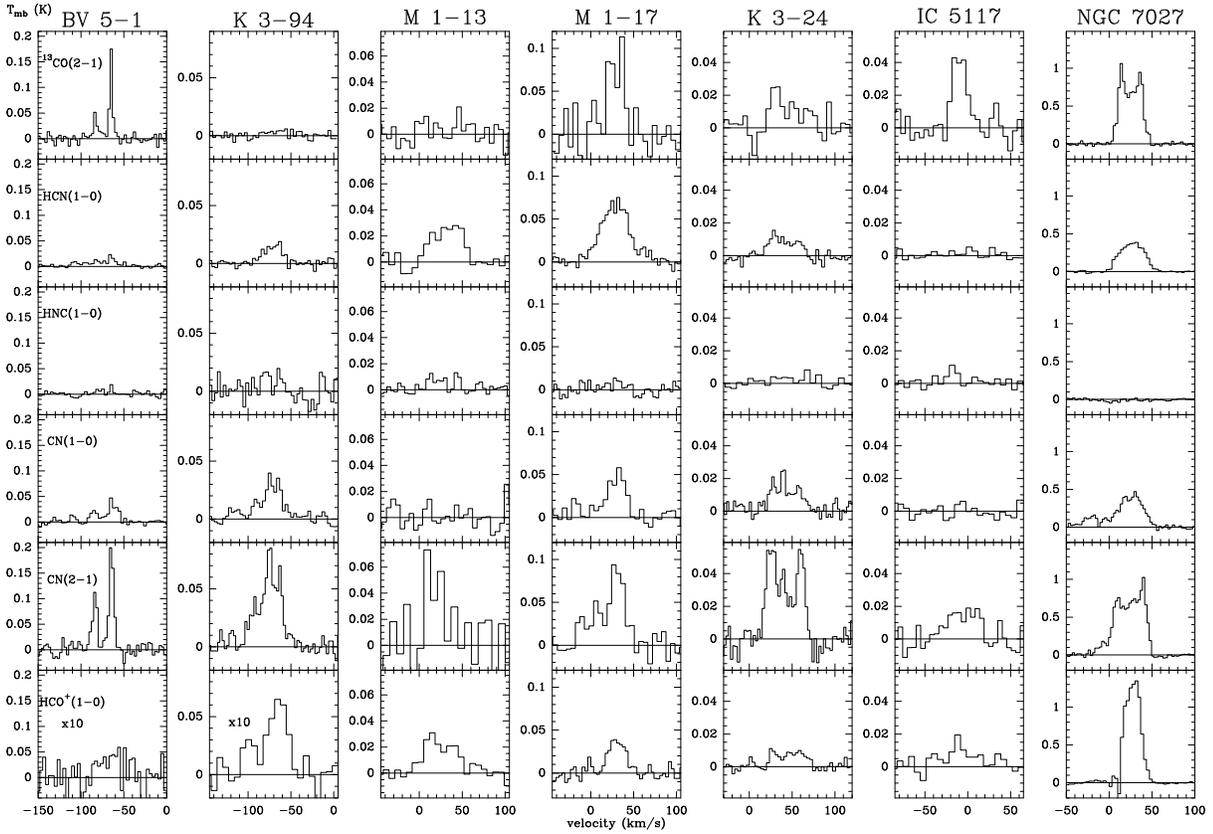,width=17cm}}
\caption[ ]{Spectra of the lines observed in the PN sample. The temperature 
scale is the same for every line for each PN, except for the HCO$^+$ line 
in PNe BV~5-1 and K~3-94, where the data have been expanded by a factor of 10. 
The velocity scale of the CN profiles are calculated with respect to 
the frequency of the strongest component of both lines 
($\nu_0 \, = \, 113.490945$~GHz for the CN(1--0) line and 
$\nu_0 \, = \, 226.874766$~GHz for the CN(2--1) line). 
Some of the data have been smoothed for display purposes in the present figure 
(line intensities reported in Table 2 have been measured at the original 
resolution). }
\label{spec}
\end{figure*}

Spectra obtained toward the sample of PNe 
are shown in Fig. \ref {spec}. Table \ref{res} lists the line 
integrated intensities and upper limits derived from such spectra. 
The integrated intensities for NGC 7027 are generally found to be compatible 
with those previously reported by Bachiller et al. (1997a). 
A significant difference was only found for the $^{13}$CO(2--1) line, 
for which we report an intensity higher by a factor of 30\%. 
We caution that slight pointing errors can result in important differences 
in the line intensities of such compact objects, and we believe that the 
observations reported here were made with better pointing accuracy.

HCO$^+$ is detected in all the PNe of our sample. At least one of the 
N-bearing molecules we looked for (HCN, HNC, CN) is also detected in 
every PN, the three molecules being detected in K 3-94 and M 1-17. 
The HNC(1--0) emission is weak, and we only have at best tentative 
detections. $^{13}$CO is detected in BV~5--1, M 1--17, K~3--24, IC~5117 
and NGC~7027. 

\begin{figure}
\centerline{\psfig{angle=-90,figure=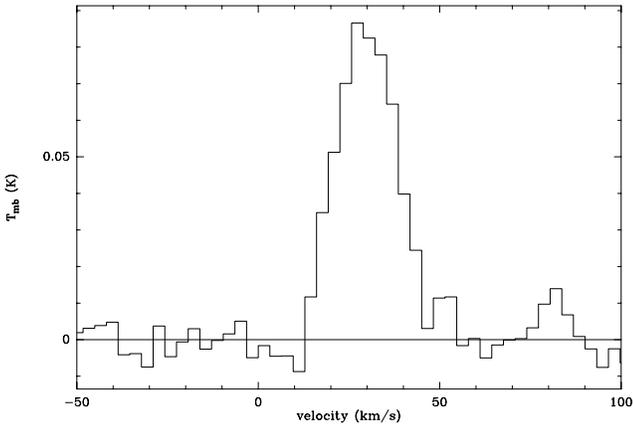,width=9cm}}
\caption[ ]{Spectrum of the N$_2$H$^+$(1--0) line towards NGC 7027. 
}
\label{n2h}
\end{figure}

We also observed the N$_2$H$^+$(1--0) line in the sample.  The first detection 
of this line in a circumstellar envelope was reported by Cox et al. (1993)
in NGC\,7027 (see their Table 1). A high-sensitivity spectrum taken toward 
NGC\,7027 is shown in Fig. \ref{n2h}, and its intensity is found to be 
consistent with that reported by Cox et al. (1993). This line was 
not detected in any other PN. The detection limits are reported 
in Table \ref{res}.

\begin{table*}
\caption[ ]{Integrated intensities of the observed lines, 
in main-beam scale. Intensities are in K km s$^{-1}$ and 
1 rms errors are given in parenthesis. The quoted values for the CN lines 
refer to the high frequency fine-structure groups $J=3/2\rightarrow 1/2$ 
for the CN(1--0) line and $J=5/2\rightarrow 3/2$ for the CN(2--1) line. 
}
\label{res}
\begin{flushleft}
\scriptsize
\begin{tabular}{lcccccccccccccc}
\hline\noalign{\smallskip}
Name   & \multicolumn{2}{c}{$^{13}$CO(2--1)}  & 
         \multicolumn{2}{c}{HCN(1--0)}        &
         \multicolumn{2}{c}{HNC(1--0)}        &
         \multicolumn{2}{c}{CN(N=1--0)}   &
         \multicolumn{2}{c}{CN(N=2--1)}   &
         \multicolumn{2}{c}{HCO$^+$(1--0)}    &
         \multicolumn{2}{c}{N$_2$H$^+$(1--0)} \\
\noalign{\smallskip}
\hline\noalign{\smallskip}
BV 5-1   &     ~~1.10 & (0.05) & ~0.54 & (0.02) &            & (0.03) & ~0.77 & (0.03) & ~2.22 & (0.08) & ~0.12 & (0.02) &      & (0.03) \\
K 3-94   &            & (0.15) & ~0.41 & (0.03) & $\sim$0.35 & (0.09) & ~0.96 & (0.05) & ~3.54 & (0.10) & ~0.18 & (0.02) &      & (0.04) \\
M 1-13   &            & (0.2)~ & ~1.16 & (0.07) & $\sim$0.30 & (0.07) &       & (0.2)~ & ~2.16 & (0.3)~ & ~0.95 & (0.07) &      & (0.05) \\
M 1-17   &     ~~2.49 & (0.3)~ & ~2.77 & (0.11) & $\sim$0.36 & (0.11) & ~1.39 & (0.2)~ & ~2.77 & (0.6)~ & ~1.18 & (0.08) &      & (0.2)~ \\
K 3-24   & $\sim$0.66 & (0.11) & ~0.58 & (0.02) &            & (0.02) & ~0.61 & (0.05) & ~3.58 & (0.12) & ~0.42 & (0.02) &      & (0.01) \\
IC 5117  &     ~~0.85 & (0.11) &       & (0.03) & $\sim$0.15 & (0.04) &       & (0.03) & ~0.81 & (0.08) & ~0.45 & (0.07) &      & (0.06) \\
NGC 7027 &     ~26.30 & (0.2)~ & 12.56 & (0.2)~ &            & (0.2)~ & 19.15 & (0.3)~ & 45.49 & (0.8)~ & 29.78 & (0.4)~ & 1.74 & (0.06) \\
\noalign{\smallskip}
\hline
\noalign{\smallskip}
\end{tabular}
\end{flushleft}
\end{table*}

\subsection{Column densities and their evolution}
Estimates of the column densities have been made under the usual 
assumptions of LTE and optically thin emission. The excitation temperature
was assumed to be uniform and equal to 25~K, which is the kinetic
temperature deduced from CO observations in a few other nebulae
(see Bachiller et al. 1993).   
When the line was not detected, we used three times the upper limits to 
the integrated intensities given at a 1$\sigma$ level in Table \ref{res} 
to derive upper limits in the corresponding column densities.  The CN 
column densities have been calculated from the CN(2--1) 
line intensities, since this line is expected to be optically thinner 
than the CN(1--0) line and, in addition, the 2--1 observations provide
higher resolution. It should be pointed out that 
the hyperfine structure of the CN spectrum may introduce some additional
uncertainties in the estimates of column densities. This is because 
of the possible anomalous excitation through 
near-infrared pumping which can produce weak masers (Bachiller et al. 1997b). 
Unfortunately, the separation of the different hyperfine lines is 
very difficult in the case of our nebulae because of the complex kinematic 
structure of the line profiles, which often include 2 or 3 components at 
close velocities (see for instance the CO profiles in Huggins 
et al. 1996 and Josselin et al 2000). We thus simply assumed that the
hyperfine line components within a fine group presented standard LTE ratios.
The resulting column density estimates
for all observed molecules are given in Table \ref{res2}. 

\begin{table*}
\caption[ ]{Column densities (in cm$^{-2}$) derived from observed line 
intensities given in Table \ref{res} (see Sect. 4.1 for details). 
}
\label{res2}
\begin{flushleft}
\begin{tabular}{lrrrrr}
\hline\noalign{\smallskip}
Name     & N($^{13}$CO)         & N(HCN)               & N(HNC)               & N(CN)                & N(HCO$^+$) \\
\noalign{\smallskip}
\hline\noalign{\smallskip}
BV 5-1   &     ~~~3.9~10$^{14}$ &     ~~~1.3~10$^{12}$ & $\less$2.4~10$^{11}$ &     ~~~3.0~10$^{12}$ & 1.7~10$^{11}$ \\
K 3-94   & $\less$1.6~10$^{14}$ &     ~~~1.0~10$^{12}$ & $\sim$ 9.1~10$^{11}$ &     ~~~4.7~10$^{12}$ & 2.6~10$^{11}$ \\
M 1-13   & $\less$2.1~10$^{14}$ &     ~~~2.9~10$^{12}$ & $\sim$ 7.9~10$^{11}$ &     ~~~2.9~10$^{12}$ & 1.4~10$^{12}$ \\
M 1-17   &     ~~~8.9~10$^{14}$ &     ~~~6.9~10$^{12}$ & $\sim$ 9.4~10$^{11}$ &     ~~~3.7~10$^{12}$ & 1.7~10$^{12}$ \\
K 3-24   &  $\sim$2.4~10$^{13}$ &     ~~~1.4~10$^{12}$ & $\less$4.2~10$^{11}$ &     ~~~4.8~10$^{12}$ & 6.0~10$^{11}$ \\
IC 5117  &     ~~~3.0~10$^{14}$ & $\less$2.2~10$^{11}$ & $\sim$ 3.8~10$^{11}$ &     ~~~1.1~10$^{12}$ & 5.5~10$^{11}$ \\
NGC 7027 &     ~~~9.4~10$^{15}$ &     ~~~3.1~10$^{13}$ & $\less$9.3~10$^{12}$ &     ~~~6.1~10$^{13}$ & 4.3~10$^{13}$ \\
\noalign{\smallskip}
\hline
\noalign{\smallskip}
\end{tabular}
\end{flushleft}
\end{table*}

For all the observed species, in all objects other than NGC\,7027, 
the column densities are relatively small which is an indication 
that the hypothesis of optically thin emission is probably appropriate. 
Even in NGC\,7027, the large column densities may not be incompatible 
with optically thin emission (Hasegawa \& Kwok 2001). However, the 
assumptions of LTE and uniform excitation temperatures are much more 
questionable. In fact, all the observed species but $^{13}$CO have 
high dipole moments ($\more$~1.5~Debye) and are thus expected to present 
sub-thermal populations.

\begin{figure}[hbt]
\centerline{\psfig{figure=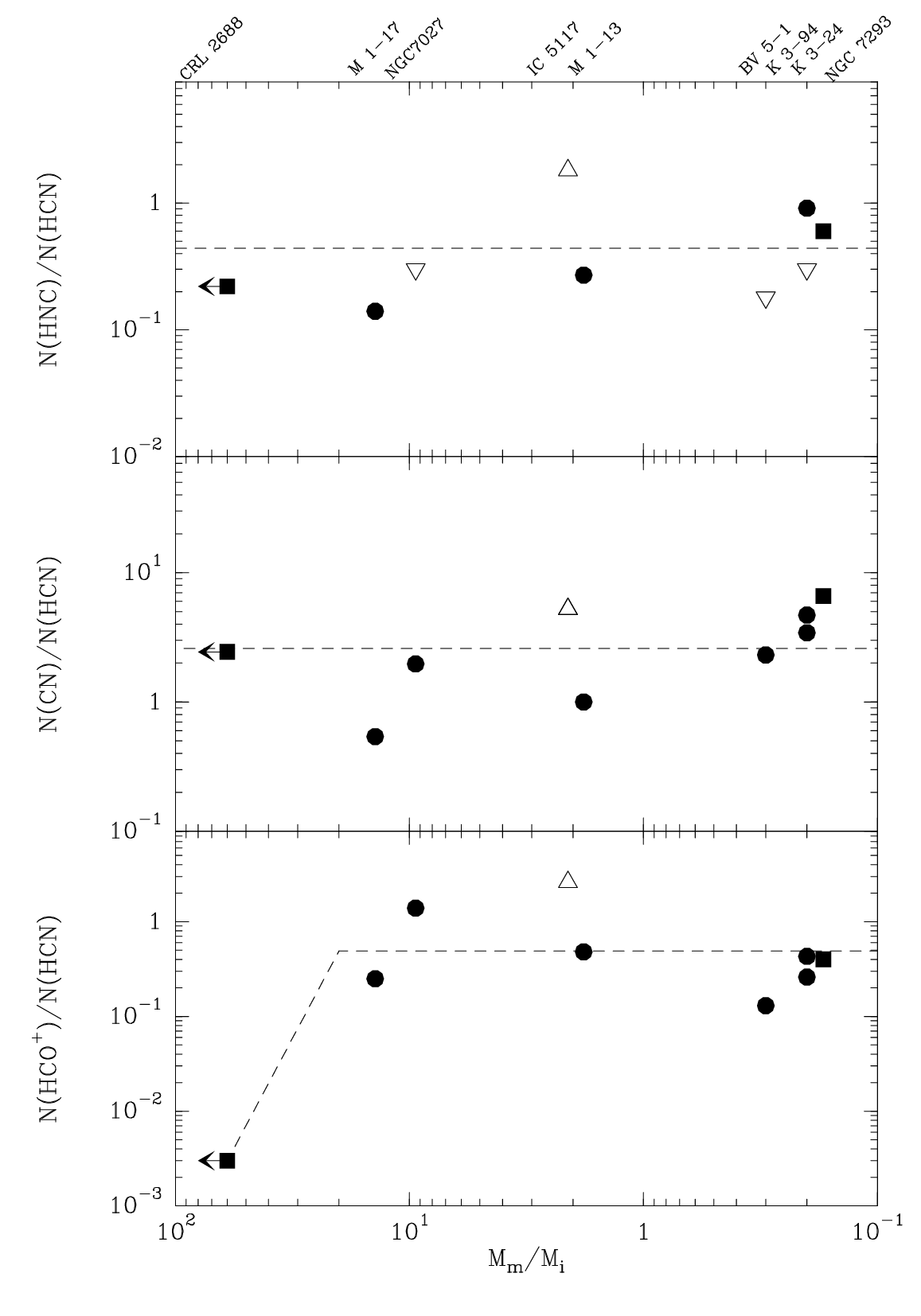,width=9cm}}
\caption[ ]{Abundances of HNC, CN and HCO$^+$ relative to HCN as a function 
of the molecular to ionized mass ratio. The ratio is plotted with a 
filled circle when both molecules are detected and with an empty triangle if 
one of the molecules is not detected, the orientation of the triangle 
indicating if the limit is a lower or an upper one. Data for CRL~2688 and 
NGC~7293, indicated by squares, are from Bachiller et al. (1997a). 
Horizontal dashed lines indicate the average of the relative abundances 
within our sample. }
\label{ratios}
\end{figure}

In order to study the evolution of the abundances in the different nebulae, 
we thus decided to use ratios with respect to HCN (which was detected in 
all nebulae but IC\,5117). Since HNC, CN and HCO$^+$ have high dipole 
moments similar to that of HCN, non-LTE effects on the column densities 
should be of the same order of magnitude for these species
so the column density ratios should be less affected by 
these uncertainties. Furthermore, Hasegawa \& Kwok (2001) found that HCN, 
CN and HCO$^+$ emission regions have similar sizes in the case of NGC\,7027, 
while CO emission is more spatially extended. Beam dilution effects should 
thus be less important when comparing emission from the observed species 
but CO. Figure \ref{ratios} 
displays the values of the HNC, CN and HCO$^+$ column densities
with respect to that of HCN. The column density ratios are plotted
as a function of the molecular to ionized mass ratio 
(M$_{\rm m}$/M$_{\rm i}$). 
Data from Bachiller et al. (1997a) for CRL~2688, a typical post-AGB star 
and the youngest object in their sample, 
and NGC~7293 (Helix, their oldest PN) have been added, in order to 
stress the evolution of the abundances. Concerning CRL~2688, one should 
keep in mind that this object is too young for ionized gas to be present, 
so its abscissa is arbitrary. Furthermore, its HCN line is optically 
thick so the relative abundances should be considered as upper limits. 

Since, as mentioned before, M$_{\rm m}$/M$_{\rm i}$  is
a good indicator of the evolutionary state of the nebula, the abscissas in
Fig. \ref{ratios} can be considered as a time axis, and the lines connecting 
the values of the column density ratios should reflect the evolution of the
abundance ratios. It is interesting to note some general trends in the 
abundance ratios. The most striking point is the dramatic increase in 
HCO$^+$ abundance relative to HCN from post-AGB stars to PNe, by more than 
two orders of magnitude. A second-order noticeable effect is the slight 
decrease of the HCO$^+$/HCN ratio from rather young PNe 
(M$_{\rm m}$/M$_{\rm i}$ $\sim$ 10) to older ones. 
The CN abundance relative to HCN is observed to increase by a factor 
of about 20 from the youngest objects in the sample (NGC\,7027 and M\,1-17) 
to the oldest ones (M$_{\rm m}$/M$_{\rm i}$ $\sim$ 0.2). We remind here that 
the rather high value found for CRL~2688 must be attributed to an 
under-estimation of the HCN column density.   
The increase in the CN abundance occurs essentially during the early stages 
of the PN evolution, the CN/HCN ratio becomes rather constant after then. 
The HNC/HCN column density ratio does not present a very clear tendency 
with age. But it should be pointed out that the HNC line was only 
tentatively detected in some of the objects, and that more sensitive
HNC observations are needed before drawing any firm conclusion. 
Nevertheless, this ratio seems again to be relatively lower in the 
two youngest objects (HNC/HCN $\less$ 0.2) than in some more evolved ones 
(HNC/HCN up to $\sim$ 1). These values are in the range of those previously 
observed in other nebulae (Bachiller et al. 1997a) and
estimated theoretically (Ali et al. 2001). 

The examination of Fig. \ref{ratios} also gives information about
the evolutionary stages of the nebulae. For instance, M~1-17, which was 
previously suspected to be a young PN because of the presence 
of wide wings in its spectrum (Bachiller et al. 1991), presents 
a high M$_{\rm m}$/M$_{\rm i}$ ratio. This,
together with its molecular abundances (see Fig. \ref{ratios}),
confirms now that M~1-17 is a very young object, similar in many respects
to NGC~7027. On the other hand,  BV~5-1, K~3-24 and K~3-94 
seem to be at a well advanced evolutionary stage, close to that of  NGC~7293. 

A remarkable exception among the observed behaviours is the case of IC~5117, 
the only nebula not detected the in HCN 1--0 line, and with rather weak 
CN emission (only the CN 2--1 line was detected), which seems typical 
of an oxygen-rich chemistry, whereas ionized gas in IC~5117 appears 
carbon-rich (C/O$\sim$1.3, Hyung et al. 2001). A C/O ratio $>$1 is 
consistent with the idea that massive PNe as those in our sample 
result from the evolution of stars with intermediate masses 
(M $\more$ 3 M$_\odot$) and thus have experienced the 3$^{\mbox{rd}}$ 
dredge-up, which makes them carbon-rich.  
However the determination of 
the carbon abundance in IC~5117 is very uncertain. Furthermore, 
the existence of massive oxygen-rich PNe is not excluded, because 
of hot-bottom burning which could convert $^{12}$C into $^{13}$C and $^{14}$N, 
decreasing the relative abundance of carbon (Renzini \& Voli 1981). 
One could also think about a chemical stratification in the nebula, as the 
molecular gas corresponds to mass ejection prior to that traced by ionized 
gas. Additional observations are needed to understand this peculiar PN. 

\subsection{Carbon isotopic ratio}
In the case of PNe, the observed I($^{12}$CO(2--1))/I($^{13}$CO(2--1)) 
intensity ratio is expected to be
a good indicator of the isotopic $^{12}$C/$^{13}$C abundance ratio
(see Palla et al. 2000). In fact, 
the intensity ratio equals the abundance ratio when the following 
assumptions are accomplished : (i) the $^{12}$CO(2--1) and $^{13}$CO(2--1) 
rotational levels are thermalized, (ii) the $^{12}$CO(2--1) 
and $^{13}$CO(2--1) are not very optically thick, and (iii) 
the beam filling factor is similar for both lines.
Since the dipole moment of CO is small 
($\sim$ 0.11 Debye), the first assumption is very plausible. In addition,
saturation effects are expected to be negligible for 
most evolved PNe, since the I(CO(2--1))/I(CO(1--0)) ratio suggests optically 
thin emission (Huggins et al. 1996, Josselin et al. 2000). But this may 
not the case for the youngest PNe (such as  NGC 7027 and M~1-17), for which 
the $^{12}$CO lines are probably optically thick; isotopic ratios for 
these objects should thus been considered as lower limits. 

The validity of the hypothesis of equal beam filling depends directly 
on the spatial distribution of the CO isotopomers,
and thus on selective processes of formation and 
destruction of the molecules. Two types of reactions may be important 
to determine the CO isotopomer ratio: selective photodissociation, $^{13}$CO
being more easily photodissociated, and chemical fractionation through the
reaction
\begin{equation}
 ^{12}CO + ^{13}C^+ \, \rightarrow \, ^{13}CO +^{12}C^+
\end{equation}
But these effects are generally thought to compensate each other
(see Mamon et al. 1988 for a more detailed discussion). 
In summary, we believe that for most of the PNe discussed here, 
$^{12}$C/$^{13}$C~=~I($^{12}$CO(2--1))/I($^{13}$CO(2--1)). The 
isotopomer ratios resulting from the observations
are summarized in Table \ref{13co}. 

\begin{table}
\caption[ ]{Carbon isotopic ratios in the PNe of our sample. 
}
\label{13co}
\begin{flushleft}
\begin{tabular}{lr}
\hline\noalign{\smallskip}
Name   & $^{12}$C/$^{13}$C  \\
       &                    \\
\noalign{\smallskip}
\hline\noalign{\smallskip}
BV 5-1   &       ~~~9 \\
K 3-94   & $\more$ 31 \\
M 1-13   & $\more$ 37 \\
M 1-17   &     ~~~ 27 \\
K 3-24   &     ~~~ 35 \\
IC 5117  &     ~~~ 23 \\
NGC 7027 &     ~~~ 11 \\
\noalign{\smallskip}
\hline
\noalign{\smallskip}
\end{tabular}
\end{flushleft}
\end{table}

We find ratio values in the  range 
$10 \less ^{12}{\rm CO}/^{13}{\rm CO} \less 40$.
The $^{12}$C/$^{13}$C ratio was recently measured in some additional
PNe by Palla et al. (2000) and Balser et al. (2002), 
and the values reported by these authors 
are in general agreement with the ones reported here. These values 
are below those expected from standard asymptotic giant branch models 
and suggest non-standard mixing processes. 

We have estimated the mass of the progenitors of BV 5-1 and K 3-94, 
by assuming relevant stellar parameters taken from the 
Acker et al. (1992) catalogue, and by following the procedure 
described in Palla et al. (2000). For M 1-17, IC 5117 and NGC 7027, 
estimates are provided by Palla et al. (2000). For M 1-13 and K 3-24,
the lack of the central star photometry prevents any realistic estimate
to be done. Of the studied sample, 
BV 5-1 and K 3-94 appear to have the most massive progenitors, with $\sim$5.5 
and $\sim$4.0 M$_\odot$, respectively (we note however 
that the estimate of the progenitor mass requires the previous estimate of 
the luminosity of the central star which in turn depends on the choice of the 
distance, which as discussed above remains quite uncertain). 
Then, when reported in Fig. 3 of Palla et al., the $^{12}$C/$^{13}$C isotopic 
ratio appears consistent with the models they quote. We thus confirm that 
non-standard mixing processes should thus be important only in low-mass 
stars, as suggested by Charbonnel (1995). 

\section{Discussion}
 
We next discuss the chemical processes which must be relevant 
in the case of the molecules observed here by 
strengthening the differences between the conditions prevailing in PNe from
those of the interstellar medium.
Special attention is devoted to the clumpiness of the neutral PN envelopes
since clumping is expected to play a key role in the preservation and 
evolution of the molecular species.

\subsection{The survival of molecules in PNe}
The compact PNe discussed here sample the evolution of the 
chemistry from the younger objects to rather developed PNe. An 
interesting result from our observations is the continuity 
of the molecular content found
for PNe with very different evolutionary status and morphologies. 
In particular, N-bearing molecules are present at all the 
sampled evolutionary stages. This finding is in contrast  
with the recent suggestion made by Hasegawa et al. (2000) that 
most molecules from the AGB remnant (except H$_2$ and CO) are completely 
destroyed and then re-formed at the PN stage. Our observations suggest 
on the contrary that most molecular species survive during the transition 
from the AGB to the PN phase. 

An exception to this continuity in the molecular content is provided by 
the sudden increase in the HCO$^+$ abundance. This ion is expected to
form in warm PDR-like regions such as the interface between ionized 
and molecular gas. 
Even if currently available models fail in reproducing the observed 
abundance, they clearly show that the abundance is sensitive to the 
total density (see e.g. Table 3 in Hasegawa \& Kwok 2001). 
Additional observational probes of these PDR-like regions would be 
provided by measuring the CO$^+$ and N$_2$H$^+$ abundances, but the
corresponding observations would require higher sensitivity than 
that of the observations presented here. 

\subsection{Chemistry of N-bearing molecules}
The chemistry of N-bearing molecules has been recently studied 
by Ali et al. (2001) with a detailed time-dependent chemical model  
suited to the conditions of evolved PNe, and by Hasegawa et al. (2000) 
and Hasegawa \& Kwok (2001) for the particular case of NGC\,7027. 
The CN/HCN ratio calculated by
Hasegawa et al. (2000) is in the range 1.4 to 6.6, depending on the
volume hydrogen density and the kinetic temperature, 
in good agreement with our average value ($\sim$2.6). 
The abundances of HCN and HNC calculated by Ali et al. (2001)
are in general agreement with the observations reported here.
However their abundance  of CN is too high by a factor of $\sim$3. 

The chemistries of the three observed molecules 
(HCN, HNC and CN) are intimately interconnected. 
CN is mainly produced through the photodissociation of 
HCN or HNC, both having similar unshielded rates ($\sim$ 1.4 10$^{-9}$ 
s$^{-1}$, Le Teuff et al. 2000). It is thus expected that CN is
mainly located in a kind of shell 
surrounding the HCN/HNC regions, where the ambient UV radiation 
coming from the central star can penetrate to photodissociate HCN, HNC,
and other related compounds. The CN and HCN emitting regions are thus
not expected to be coincident, and the CN/HCN abundance ratio can 
strongly vary with the position in the nebula. The CN/HCN column
density ratios provided above must be considered as average values.

One could ask about the potential importance of two body reactions
in the chemistry of nitrogen. Reactions such as that of atomic 
carbon with NO and N$_2$ or atomic nitrogen with CH or CS, should 
have negligible contributions, both because such reactions have comparatively 
lower rates and the reactants must have very low abundances. For example, 
atomic carbon is only present in a very thin transition region between 
molecular gas (CO) and ionized gas (C$^+$), as carbon ionization potential 
and CO dissociation energy are similar ($\sim$ 11 eV). 

One exception may be the reaction 
\begin{equation}  
O + HCN \, \rightarrow \, CN + OH 
\end{equation}
as atomic oxygen may be abundant in photodissociation regions. 
Unfortunately, the abundance of atomic oxygen is not well known in these
regions, and it will certainly strongly depend on the carbon-rich or
oxygen-rich character of the envelope.
If atomic oxygen turned out to be significantly abundant, 
this reaction could accentuate the production of CN at the expense 
of HCN. Nevertheless, the presence of atomic oxygen 
can also contribute to the destruction of CN:
\begin{equation}  
O + CN \, \rightarrow \, CO + N \; \mbox{or} \; NO + C 
\end{equation}
The relative importance of these processes should be examined in the 
peculiar case of PN IC~5117 (see Sect. 3.2). 

In principle,
HCN and HNC can convert to each other through hydrogen exchange reaction 
\begin{equation}  
H + HNC \, \leftrightarrow \, HCN + H
\end{equation}
However, the forward reaction must be predominant since it has an activation 
energy four times smaller than the reverse reaction 
(Talbi et al. 1996). On the other hand, 
the HNC/HCN ratio is observed to decrease at increasingly high densities 
in interstellar clouds (Turner et al. 1997).
The relatively low HNC/HCN ratio observed in PNe could then be due to
the combination of moderate temperatures and densities prevailing in the 
molecular envelopes.
 
HCN and HNC can indeed react with several molecular ions such as HCO$^+$, 
H$_3^+$, H$_3$O$^+$ or C$_2$H$^+$ to form HCNH$^+$, which after dissociative 
recombination will lead to CN, HCN or HNC in a ratio 2/1/1 (Le Teuff et 
al. 2000). As the abundance of molecular ions is expected to increase 
with the state of evolution (as the nebula is being progressively
ionized), such reactions are expected to additionally enhance the CN/HCN 
ratio as the evolution proceeds. 

\subsection{Molecular ions: HCO$^+$ and N$_2$H$^+$}
As already discussed by Bachiller et al. (1997a), a remarkable result 
concerning the chemical content of PNe is the rather high column densities
of HCO$^+$. The ``standard'' path of  HCO$^+$ formation which may 
dominate in dense 
interstellar clouds invokes the reaction of H$_3^+$ with CO, and thus 
requires a high formation rate of H$_3^+$ through cosmic-ray ionization 
of H$_2$ (Turner 1995). However the survival 
of H$_3^+$ ion is expected to be very short, because of its very effective 
dissociative  recombination (Bachiller et al. 1997a), so this process may 
not be efficient enough to explain the observed high abundances. 
It has been pointed out by Ali et al. (2001) that
X-ray emission from the central hot star of PNe 
may also contribute to a high rate of ionization. 
In fact, the relatively high X-ray fluxes have been observed in several PNe
(e.g.: Apparao \& Tarafdar 1989) are believed to be generated from the central
star (not from the nebula). The effect of such X-rays on the envelope
can be particularly important in the case of compact PNe as these
studied here.

An additional way to produce HCO$^+$ is through the reaction 
\begin{equation}  
CO^+ + H_2 \, \rightarrow \, HCO^+ + H
\end{equation}
CO$^+$ could be formed through charge transfer between H$^+$ and CO, via 
the reaction 
\begin{equation}  
C^+ + OH \, \rightarrow \, CO^+ + H
\end{equation}
or by the direct ionization of CO by the means, for instance, of the
propagation of an ionizing front into the molecular gas of high CO abundance 
(Latter et al. 1993). The latter path may be dominant at least in some PNe. 
The initiation of such ionizing front
is one of the characteristics of PPNe (Kwok 1993) and HCO$^+$ seems
indeed to form essentially during this stage (Bachiller et al. 1997a). 
Besides the formation of HCO$^+$, the other main destruction process 
of CO$^+$ is dissociative recombination. A column density 
$N({\rm CO}^+) \, \more \, 10^{12}$~cm$^{-2}$ can be expected and is 
consistent with the detection in NGC 7027 (Latter et al. 1993). A 
systematic search for this ion in PNe where HCO$^+$ emission is intense 
would confirm this scenario.

HCO$^+$ is mainly destroyed by dissociative recombination with electrons.
These chemical processes are included in the detailed model of Ali et al. 
(2001) that leads to an abundance ratio HCO$^+$/HCN$\sim$0.1, in 
reasonable agreement with the less evolved PNe of our sample, but 
significantly lower (by a factor of 5 to 10) that the values found 
in PNe at intermediate stage of evolution (see also Bachiller et al. 
1997a). The observed trend in the evolution of HCO$^+$ abundance 
seems to imply an increase in electron density during the late stages 
of PN evolution. Hasegawa \& Kwok (2001) found that in NGC\,7027 about 90\% 
of carbon, which is a major source of electrons, is in the form of C$^+$. 
Carbon with higher ionization degree (C$^{2+}$ and C$^{3+}$) is commonly 
observed toward more evolved PNe. An increase in the destruction process 
of HCO$^+$ during the last evolutionary stages is thus expected and consistent 
with our observations. 
The abundance of  HCO$^+$ estimated by Hasegawa et al. (2000) for
NGC\,7027 is in general agreement with our observations, but it
assumes that it is formed in gas at a temperature of $\sim$800\,K.
Observations of spectral lines from higher J levels (at higher frequencies)
would be useful to estimate the gas physical conditions.

In molecular clouds, N$_2$H$^+$ is often thought to be associated with 
HCO$^+$, since it is essentially formed through 
\begin{equation}  
H_3^+  +  N_2 \, \rightarrow \,  N_2H^+ + H   \;\;\;\;\;\;
\end{equation}
But contrary to HCO$^+$, no important alternative formation is expected. 
The abundance of  N$_2$H$^+$ was however not reported in the models of 
Ali et al. (2001). The lack of other significant paths of formation,
together with its high molecular dipole moment ($\sim$ 3.4 Debye) explains 
why we only detected this ion in NGC~7027. Assuming similar line 
intensity ratios in the other PNe, one needs to reach a rms $\less$ 0.5 mK 
to allow its detection, far below the limit we obtained. A deeper search 
for this ion would be interesting to examine any correlation with the 
temperature of the central star and in particular with its X-ray flux
(when $T_\star \, \more \, 10^5$~K).

\subsection{Clumpiness and PDR-like regions}
Beside the evolutionary trends described above, a rather large dispersion  
of abundance ratios is observed for different values of 
M$_{\rm m}$/M$_{\rm i}$. 
Such scatter may be, at least partly, attributed to the the number 
and density of the clumps, which are expected to vary from one PN to another. 
Indeed these variations should act on the observed abundances by two ways. 
First, the degree of clumpiness (filling factor) can be different for 
different molecules in the highly fragmented envelopes of the more 
evolved objects such as BV 5--1 (Josselin \& Bachiller 2001). 
Second, the survival of the molecules is expected to strongly depend on the 
number and density of the clumps which determine the shielding of dissociating 
UV radiation from the central star. Indeed, high resolution maps of 
the molecular envelopes of evolved PNe show that the molecular material
is always confined within the 
ionized nebula (see e.g. Speck et al. 2002 for the Helix nebula and Josselin 
\& Bachiller 2001 for BV 5--1), in contradiction with the simple image of an 
initially neutral homogeneous nebula being gradually ionized (``Str\"omgren 
sphere''). As confirmed by direct observations of some individual dense
clumps in the Helix (Huggins et al. 1992, 2002), the 
molecules are in fact concentrated within dense blobs
which reach a state of approximate dynamical equilibrium with
the surrounding ionized material, and 
whose evolution is determined by their density, i.e. by their capability
to be  efficiently shielded from the ambient UV radiation. 
Any  consistent chemical model of the neutral matter in PNe should 
thus take into account this clumpy structure which seems to be
an essential ingredient of PNe.

\section{Conclusions}
The presence of molecular gas in compact PNe provides a unique opportunity 
to examine a rapidly evolving chemistry in a peculiar medium which is
subjected to strong radiation fields. The observations reported in 
this paper show that some species such as HCO$^+$ and CN are particularly 
abundant compared to envelopes around AGB stars or even to interstellar 
clouds. The chemical composition of the molecular envelopes in these
compact PNe appears somewhat intermediate between the composition
of proto-PNe (such as CRL\,2688 or CRL\,618) and well evolved PNe
(such as the Ring, M4--9, or the Helix).

The molecular abundances are well explained, in general terms, by
time-dependent gas-phase chemical models in which the ionization rate
is enhanced by several orders of magnitude with respect to the average
interstellar value (Ali et al. 2001, Hasegawa et al. 2000, 
Hasegawa \& Kwok 2001). Thus, our observations confirm that the 
chemistry in the neutral shells of PNe is essentially governed by the
high energy radiation from the hot central stars. Shocks due to fast 
winds which develop during the formation of the PN can also be important,
but mainly during the proto-PN stage and in the regions where the jets impact
on the envelope. The complexity of the chemical processes is increased by 
numerous factors linked to the properties of the central star and 
the geometry and degree of clumpiness of the envelope.
The survival of molecules can only occur in well protected dense
clumps, and a rich photo-chemistry very likely takes place at the 
surfaces of these PDR-like regions. 

Although the main processes governing the chemistry in the neutral envelopes
of PNe seem well identified, some significant discrepancies 
with the models remain to be understood.
For instance the calculated CN/HCN ratio is a factor of $\sim$5 higher
than the observed value, and the observed HCO$^+$ abundances tend to
be higher than those predicted by the models. In addition, 
as discussed by Hasegawa et al. (2000), some of the 
molecules observed in evolved PNe (such as the Ring or the Helix) 
could not be direct remnants from the AGB envelope, but
newly formed species, this is the case of HCO$^+$. 
However, the observations suggest that there is
a continuity in the chemical composition from the proto-PN phase to
the full photodissociation of the envelope, and that molecules like
CO and HCN are progressively photo-dissociated.

Additional observations are now required to push up our understanding 
of this chemistry. The search for species such as CO$^+$ would be useful 
to confirm the chemical networks invoked to explain the observed abundances. 

\begin{acknowledgements} 
The authors acknowledge many interesting discussions and collaborative
work on PNe with Drs. P. Cox, T, Forveille, and P.J. Huggins. 
RB acknowledges warm hospitality during a stay at GRAAL (University of 
Montpellier) during which part of this manuscript was written. 
EJ acknowledges partial financial support from french CNRS grant ATIPE 
and RB from spanish MCYT grant AYA2000-927.
\end{acknowledgements}

\end{document}